\documentclass[11pt,fleqn]{article}
\usepackage{amssymb}

\usepackage{amsmath}

\usepackage{amsmath,amssymb,amsthm}

\begin{document}
{\Large
\begin{center}
{\Large \textbf{Ansatzes and exact solutions for nonlinear
Schr\"odinger equations}}

\vskip 20pt {\large \textbf{Irina YEHORCHENKO}}

\vskip 20pt {\small {Institute of Mathematics of NAS Ukraine, 3
Tereshchenkivs'ka Str., 01601 Kyiv-4, Ukraine}

{\it e-mail: iyegorch@imath.kiev.ua}}
\end{center}

\vskip 50pt
\begin{abstract}We consider construction of ansatzes for nonlinear
Schr\"odinger equations in three space dimensions and arbitrary nonlinearity,
and conditions of their reduction to ordinary differential equations. Complete
description of ansatzes of certain types is presented. We also discuss the relationship between solutions, and both Lie and conditional symmetry of these equations.
\end{abstract}

\section{Introduction}

Reduction of a multi-dimensional partial differential equations (PDEs) to ordinary differential equations (ODEs) or PDEs with fewer independent variables is an efficient method for finding of exact solutions of initial PDEs. 

The two principal methods to achieve such reduction are direct substitution of some special expression into the PDE, and reduction by means of non-conjugate subalgebras of the symmetry algebra possessed by the PDE in question. Both methods use the notion of ansatz -- a special substitution which reduces 
a PDE to another PDE or to an ODE.

The direct method (presented as a systematic method in \cite{ClarksonKruskal}, though used as ad hoc method many times before) is substitution of some special expression into the initial equation. The form of such expression is often the simplest -

$$ u = \phi (\omega) $$ 

\noindent
where $u$ is the dependent function (or vector-function, or set of functions) in the original equation, $\phi$ is the new dependent function (or vector-function, or set of functions) in the reduced equation, and $\omega$ is the set of new independent variables (the number of variables in this set is smaller that the number of independent variables in the initial equation. The number of elements in the sets $u$ and $\phi$ is the same.

It is possible to consider special types of reduction when the initial equation is reduced to equation with smaller or larger number of dependent variables, but we will not consider such cases here. It is also possible to take some guessed ansatz, or a generalization of some known ansatzes.

The classical Lie method of similarity reduction of PDEs \cite{{Lie},{Ovs-eng},{Olver1}} makes use of Lie symmetry properties of the equation under study. The idea of this approach also focuses on a notion of ansatz  \cite{{Lie},{FZbirnyk81},{FS83}}. The Lie method provides ansatzes using subalgebras of an invariance algebra of an equation \cite{Lie},\cite{Ovs-eng},\cite{Olver1},\cite{FBarBook}.

Both these methods have certain advantages and drawbacks. The direct method sometimes allows obtaining a wider class of exact solutions than the Lie similarity method. However, it is less algorithmic, and the resulting reduction conditions are often extremely difficult, if possible at all, to solve. On the opposite, the Lie method is more algorithmic.

We noticed another benefit of the direct method - it allows to obtain inequivalent solutions of the initial equation of the special form, and often the Lie method gives the same reduced equations and exact solutions for non-conjugate subalgebras of the Lie symmetry algebra.

We tried to search for particular ansatzes, substituting this general form of ansatz to the equation in question, and then considering conditions of its reduction.

This technique was used intensively for various two-dimensional equations
(see. e.g. \cite{BlumanCole}-\cite{FUMZh91}).

In this paper we succeeded to apply this technique for a four-dimensional
equation. Our results can be generalised to multi-dimensional equations, and this is another advantage compared to the Lie similarity method, where the set of non-conjugate subalgebras of the Lie symmetry algebra has to be built specifically for each particular number of independent variables.

The general idea is quite obvious, but the main difficulties here are
investigation of compatibility and solution of  reduction  conditions,
which present nontrivial problems in the multi-dimensional case.

In this paper we use a combination of both methods - we apply the straightforward (direct) method with the ansatz that is the most general ansatz found in the process of the Lie similarity reduction of the nonlinear Schr\"odinger equation in the papers \cite{FS87}-\cite{GagnonWinternitz88}. 

We consider inequivalence of reductions up to equivalence algebra of the reduction conditions. The obtained non-equivalent ansatzes can be used for reduction of other equations, e.g. we showed an example of the nonlinear wave equation.

This paper is based on the work started in \cite{FY91}. Here we also corrected some mistakes from that paper.

\section{Reduction of the general nonlinear Schr\"odinger equation}

We consider the general form of the
nonlinear Schr\"odinger equation with an arbitrary function in the nonlinear part

\begin{equation} \label{Schr}
2i u_t +\triangle u-uF(|u|) =0.
\end{equation}

Here $u$ is a complex valued function, $u=u(t,\overrightarrow{x})$, $\overrightarrow{x}$ is
an $n$-dimensional vector of space variables, $|u| =\sqrt{uu^*}$,
an asterisk designates complex conjugation,
$\triangle u =\partial^2u/\partial x_a^2$, $ a=1,...,n$.

We point out once more that in the further consideration of reduction the function $F$ is arbitrary, as for particular forms of $F$ the reduction conditions may be different.

Eq. (\ref{Schr}) with an arbitrary function $F$ is invariant under the Galilei
algebra with basis  operators
\begin{gather}
\partial_t =\frac{\partial}{\partial t},\ \ \partial_a,\ \ J_{ab} =x_a\partial_b -x_b \partial_a,\nonumber \\
G_a =t\partial_a +ix_a (u\partial_u -u^* \partial_{u^*} ),\quad a,b=1,...,n,\
M=i(u\partial_u -u^* \partial_{u^*}) \label{GA}
\end{gather}

Solutions obtained from the algebra (\ref{GA}) by means of the Lie method are well-known
\cite{FS87}-\cite{GagnonWinternitz88}, and all of them are of the form

\begin{equation}
u=\exp \{ if (t,\overrightarrow{x})\}\,\varphi (\omega). \label{anz}
\end{equation}

Such ansatz is the most general substitution reducing an arbitrary nonlinear equation (\ref{Schr}) to an ODE.
The expression (\ref{anz}) where $f$, $\omega$ are some unknown real functions of $t$
and $\overrightarrow{x}$ will be an ansatz for Eq.(\ref{Schr}) if its substitution reduces (\ref{Schr}) to an ODE
for a complex function depending on the new variable $\omega$ only. 

Actually in this situation we perform absolute reduction of two differential invariants of the algebra (\ref{GA}) - $\frac{2i u_t +\triangle u}{u}$ and $|u|$ (see \cite{FY89}).

{\bf Definition 1}. We say that an ansatz $u = \phi (\omega)$ absolutely reduces an expression $\Phi(x,u,\underset{1}{u},\ldots , \underset{r}{u})$, if substitution of this ansatz into the expression gives $\Phi(\omega,\phi,\underset{1}{\phi},\ldots , \underset{r}{\phi})$ without any multipliers depending on "old" variables.

If an ansatz absolutely reduces some differential invariants, it also reduces any equations, built as expressions of these differential invariants.

If we apply the direct method to some special forms of the function $F$, we may receive also "relative reductions" - when the substitution of an ansatz will give a reduced equation with some multiplier depending on "old" variables.

\section{Reduction conditions and their general solution}

Substituting the ansatz (\ref{anz}) into the equation (\ref{Schr}), we
get conditions on the functions $f$ and $\omega$:
\begin{gather}
 2f_t +f_af_a =S(\omega ),\ \triangle f =T(\omega),\\
 \omega_t +f_a \omega_a =X(\omega), \\
  \triangle \omega =Y (\omega),\ \omega_a \omega_a =Z(\omega), \label{red-cond}
\end{gather}
where $S$, $T$, $X$, $Y$, $Z$ are arbitrary smooth functions.

For $n=2$, $n=3$ we had found the general solution of the system (\ref{red-cond}) up to equivalence of substitutions (\ref{anz}).

For the purpose of classification of reductions of Eq.(\ref{Schr}), it is  sufficient  to  consider the system (\ref{red-cond}) only up to equivalence of the ansatzes (\ref{anz}). 

We shall call ansatzes equivalent if they lead to the same solutions of the equation.

We deal here with real functions $f$ and $\omega$, so $Z(\omega)$ in (\ref{red-cond}) must be nonnegative. Whence we can reduce by local transformations the equation $\omega_a\omega_a=Z(\omega)$ to the same form with $Z(\omega) =0$ or $Z(\omega)=1$.

Below we consider separately the cases $Z(\omega) = 0$ and $Z(\omega)=1$.

1)\ {\bf $Z(\omega) =0$}. In this case $\omega_a =0$, $\omega =\omega (t)$ and
we can put $\omega =t$. The system (\ref{red-cond}) can be written as
\begin{equation}
2f_t +f_af_a =S(t),\quad \triangle f =T(t).\label{cond-f}
\end{equation}

It is evident that the ansatzes of form (\ref{anz}) are equivalent up to
transformations $f \to f+r(\omega )$, so we can put $S(t)=0$.
We come to the system

\begin{equation}
2f_t +f_af_a =0,\quad \triangle f=T(t),\label{HJ}
\end{equation}

and the following theorem gives a necessary condition of its compatibility.

{\bf Theorem 1.} \ {\it The system (\ref{cond-f}) can be compatible only if}
\noindent
$$T(t) =\theta'(t)\theta (t), \ \theta^{(n+1)} =0$$.

Proof of this theorem can be carried out using differential
consequences of (\ref{cond-f}) and the Hamilton-Cayley theorem. It is rather
cumbersome, and its complete version can be found in \cite{FY91}. The proof is based on the methods developed in \cite{FZhR90}.

2)\ {\bf $Z(\omega )=1$}. It had been established in \cite{Collins76} that when $n=3$,\\
$\triangle \omega = N /\omega$, $N=0$,1,2 ($N=0$,$1$ for $n=2$). Up to
equivalence of ansatzes, we can put $X(\omega)=0$.

{\bf Theorem 2.} \ {\it The system of equations
\begin{equation}
2f_t +f_af_a =S(\omega), \ \triangle f =T(\omega),
f_a\omega_a +\omega_t =0, \ \omega_a \omega_a =1, \ \triangle \omega
=N /\omega, \label{red-cond}
\end{equation}
where $N=0,1$ with $n=2$, $N=0,\,1,\,2$ with $n=3$ is compatible iff

\noindent
$T(\omega) =0$; \ $S(\omega) =c_1 \omega +c_2$, \ $N=0$;
$S(\omega) =c_1 \ \omega ^2 +c_2$,\ $ N=1$; 

\noindent
$ S(\omega )=c_1$, \ $ N=2$; \
$c_1,\ c_2$ are arbitrary constants. }

{\bf Theorem 3.} \ {\it The system ({\rm \ref{red-cond}}) is invariant with respect to the operators}
\begin{equation}
\partial_a,\ \ \ J_{ab}=x_a\partial_b -x_b \partial_a,\ \ \ \widehat{G}_a=t\partial_a +x_a \partial_f.
                  \label {IA-cond}
\end{equation}

Thus, we can search for its general solution up to transformations generated
by operators (\ref{IA-cond}):
\begin{equation}
x_a\to \alpha_{ab} x_b +\beta _a,\quad x_a \to g_a t +x_a, \label{transf}
\end{equation}

$\alpha_{ab},\ \beta_a,\ g_a$ are constants, $\alpha_{ac} \alpha_{cb} =\delta_{ab}$
(the Kronecker symbol).

Further we adduce all solutions of the system (\ref{red-cond}), which are inequivalent up
to transformations (\ref{transf}).

{\bf Theorem 4.} {\it All solutions of the system} (\ref{red-cond}) {\it inequivalent under up
to transformations }(\ref{transf})  {\it can be represented by solutions in the following lists}:
\noindent
{\bf I.}\ $Z(\omega)=0$, $\omega =t$:

1)$n=3$,
\begin{equation}
f=\frac{1}{2}\{\frac{x_1^2}{t+A_1}+\frac{x_2^2}{t+A_2}+\frac{x_3^2}{t+A_3}\}; \label{n3}
\end{equation}

2)$n=2,3$, 

$$f=\frac{1}{2}\{\frac{x_1^2}{t+B_1}+\frac{x_2^2}{t+B_2}\};$$

3)$n=2,3$, 

$$f=\frac{x_1^2}{2t +c_1};$$

4)$n=2,3$, 

$$f=c_2 x_1 +c_3-\frac{1}{2} c_2^2 t.$$

\noindent
{\bf II.}\ $Z(\omega)=1$:

1)$n=2$,3, 
$$\omega =x_1 +at^2, \ f=-2at x_1 -\frac{4}{3}a^2 t^3 +bt;$$

2)$n=2,3$, 

$$\omega =(x_1^2 +x_2^2)^\frac{1}{2}, \ f=c\,\tan^{-1} \frac{x_1}{x_2}+\alpha t;$$

3)$n=3$, 

$$\omega =(x_1^2 +x_2^2 +x_3^2)^\frac{1}{2},\ f=\beta t.$$

Here $A_i$, $B_i$, $C_i$, $a$, $b$, $c$,  $\alpha$, $\beta$ are arbitrary constants.

The ansatz (\ref{anz}) reduces Eq.(\ref{Schr}) to the following ODE:

\begin{equation}
-2S(\omega) \varphi +iT (\omega) \varphi +2i X(\omega) \dot\varphi +
      Y(\omega)\dot\varphi+ Z(\omega) \ddot\varphi =\varphi F (|\varphi|). \label{red-eq}
\end{equation}

It follows from compatibility conditions of the system (\ref{red-cond}) that two types
of Eq.(\ref{red-eq}) are possible:

\noindent
1) If \ $\omega_a \omega_a =Z(\omega) =0$, we take $\omega =t$, and Eq.(\ref{red-eq})
will be of the form
\begin{equation}
i(2\dot\varphi +T(t) \varphi) = \varphi F(|\varphi |),\label{red-eq1}
\end{equation}
where $T=\sum\limits_{i=1}^{m}\frac{1}{t+B_i}$, $m$ may take values from
$1$ to $n$; or $T=0$.

\noindent
Eq.(\ref{red-eq1}) can be easily solved in quadratures:
\begin{equation}
{\rm if}\ T\ne 0,\ \ \varphi =r\exp \frac{i}{2} \{ \sum\limits_{i=1}^{m}\frac{x_l^2}{t+B_l} -\int F(r) dt \},\ \ r=C[(t+B_1)... (t+B_m)]^{1/2}
\end{equation}
or if \ $T=0,\ f=c_1 x_1 +c_2 -\frac{1}{2} c_1^2 t$ then
$
\varphi =c\exp i \left \{ c_1 x_1 -\frac{1}{2} F(c) t +c_2 -c_1^2 \frac{t}{2} \right\}.
$
\noindent
2) If \ $\omega_a \omega_a =Z(\omega) =1$ then Eq.(\ref{red-eq}) will be of the form
\begin{equation}
-2S(\omega) \varphi +\frac{N}{\omega} \dot \varphi +\ddot\varphi =\varphi F(|\varphi |).\label{red-eq2}
\end{equation}

Eq.(\ref{red-eq2}) in general obviously cannot be solved in quadratures. Some of its
particular  solutions were given in \cite{FS87}-\cite{GagnonWinternitz88}.

All the above results can be generalized to the case of $n$ space variables with arbitrary $n$ - the same types of ansatzes reduce Eq.(\ref{Schr}) to ODEs. It is obviously more difficult to prove that such ansatzes will exhaust the list of ansatzes that reduce the initial equation up to equivalence.

\section{Example of application of results: the nonlinear wave equation}

We can apply the results for the Schr\"odinger equation
(\ref{Schr}) to describe all inequivalent ansatzes of the form

\begin{equation}
u=f(x) \varphi (\omega) \label {anz-wave}
\end{equation}
with satisfying the conditions $\omega =\alpha_{\mu} x_{\mu},\ \alpha_{\mu} \alpha_{\mu} =0$

for a nonlinear wave equation
\begin{equation}
\Box u =\lambda u^k,  \label{eq-wave}
\end{equation}

\noindent
where $u=u(x_0,x_1,x_2,x_3)$ is a real function; $k\ne 1$, $\lambda$ are
parameters; the summation over repeated Greek indices is as follows:
$x_{\mu} x_{\mu} \equiv x_0^2 -x_1^2 -x_2^2 -x_3^2$.

Further for simplicity of presentation we shall take $\omega =x_0 +x_3$.
In this case the ansatz (\ref{anz-wave}) will reduce Eq.(\ref{eq-wave}) to an ODE if
$f(x)$ satisfies the following conditions:
\begin{equation}
\Box f =f^k T(\omega), \  2 (f_0 - f_3) =f^k Y(\omega).  \label{red-cond-wave}
\end{equation}

Here $Y(\omega)$ must not vanish. By means of a substitution of the
form $f\to \gamma (\omega) f$ (ansatzes (\ref{anz-wave}) are equivalent up to such
substitutions) we can get the system (\ref{red-cond-wave}) with $Y=\frac{2}{1-k}$.
Then from the second equation of (\ref{red-cond-wave}) we get
\begin{equation}
f=\left[ \Phi (\omega,x_1, x_2) +\frac{1}{2} (x_0 -x_3) \right]^\frac{1}{1-k}.
                                \label{red-cond-wave2}
\end{equation}
Substitution of (\ref{red-cond-wave2}) into the first equation of (\ref{red-cond-wave}) gives the
following system for the function $\Phi$:
\begin{equation}
\Phi_{11} +\Phi_{22} =T(\omega ) (1-k), \
2\Phi_{\omega} -\Phi_1^2 -\Phi_2^2 =0.
\end{equation}

Using the results for the system (\ref{red-cond}), we get solutions for different
$T(\omega)$ with which the system (\ref{red-cond-wave}) can be compatible:
\begin{gather}
\Phi =-\frac{1}{2} \sum\limits_{i=1}^{m}\frac{x_i^2}{\omega +B_i},\quad
T=\frac{1}{k-1} \sum\limits_{i=1}^{m}\frac{1}{\omega +B_i};\quad
(m=1\ \mbox {or}\ 2)\\
\Phi =B_1 x_1 +B_2 +\frac{B_1^2}{2} \omega,\quad T=0; \
B_i \ {\rm are \ constants.}
\end{gather}

Now Eq.(\ref{eq-wave}) can be reduced to the ODE
$$\varphi'\frac{2}{1-k} +T(\omega) \varphi =\lambda \varphi^k,$$
which is solvable in quadratures: e.g. let $T=\frac{1}{ k-1} \sum\limits_{i=1}^{2}
\frac{1}{\omega +B_i}$. Then

\begin{equation}
\varphi =\sqrt \rho \left[ {\lambda (1-k)^2 }{ 2}  \int \rho^{\frac{k-1}{2}}
d\omega \right]^{\frac{1}{1-k}},\quad \rho =(\omega +B_1)(\omega +B_2).
\end{equation}

These results can be easily generalized for the cases when $\omega$ is a
solution of the system $\Box \omega =0,\ \omega_{\mu} \omega_{\mu} =0$
(see e.g. \cite{FZhR90}) or when 
\noindent
$u=u(x_0,x_1,...,x_n),\ n> 3$.

Reduction and solutions for Eq.(\ref{eq-wave}) when $u$ is a complex function are
considered in \cite{FY89}.

\section{Relation between symmetry and reduction of partial differential equations.}
In general, an ansatz which reduces a PDE to another PDE with fewer independent
variables or to an ODE corresponds to some $Q$-conditional symmetry of
that equation \cite{zhdanov&tsyfra&popovych99}. 

The notion of conditional symmetry was introduced in \cite{FNBook87},
and many examples of such symmetries for various equations are given in \cite{BlumanCole}-\cite{FUMZh91}.

{\bf Definition 2}. Let us consider a PDE
\begin{equation}
\Phi(x_1,u,\underset{1}{u},\ldots , \underset{r}{u})=0 \label{F}
\end{equation}
where $x$ is a vector of independent variables, $U$ is some function,
$\underset{k}{u} $ is a set of $k$-th order partial derivatives. We shall say that
Eq.(\ref{F}) is $Q$-conditionally invariant with respect to a set of operators in involution
$$\{Q_a =\xi^{ab} (x,u) \partial_b +\eta^a (x,u) \partial_u \}$$
if the system containing Eq.(\ref{F}) and the additional conditions
\begin{equation}
L_a =\xi ^{ab} (x,u) u_b -\eta^a (x,u) =0  \label{add-cond}
\end{equation}
is compatible and invariant with respect to these operators.

Operators of conditional invariance can be defined up to an arbitrary
multiplier,and such invariance is essential when $Q_a$  are not proportional
to some operators of Lie invariance.

In the case of $Q$-conditional invariance a solution
of the system (\ref{add-cond}) gives an ansatz which will
reduce Eq.(\ref{F}). Very often investigation of reduction conditions
or $Q$-conditional invariance gives more ansatzes than the classical
Lie method. However, all ansatzes described above, correspond to Lie
symmetry operators of Eqs (\ref{Schr}). To prove this statement, it is sufficient to calculate reduction operators corresponding to ansatzes listed above, and see that they are equivalent to Lie symmetry operators.

\section{Conclusions}

In this paper we presented all inequivalent ansatzes of the form (\ref{anz}). 
The ansatzes of the form (\ref{anz}) giving reduction of the general nonlinear Schr\"odinger equation yield no essential $Q$-conditional invariance for such equation with the arbitrary $F$.

This fact does not disprove the idea that the direct method of reduction may provide more general than the classical Lie method, though it is usually more difficult to apply. 

$Q$-conditional invariance for the equation (\ref{Schr}) with special cases of the function $F$ was studied in \cite{FCh}, and for (\ref{eq-wave}) - in \cite{{BarMosk}, {zhdanov-panchakBoxu}}.

\end{document}